\documentclass[twocolumn,nofootinbib,prl,floatfix,preprintnumbers,amsmath,amssymb,groupedaddress,superscriptaddress]{revtex4}
\usepackage{dsfont}
\usepackage{graphicx} 
\usepackage{epstopdf} 
\usepackage{slashed}

\newcommand{\GeV}{\,\mathrm{GeV}}

\newcommand{\genericT}{{\ensuremath{\rm T}}}

\newcommand{\mptvec}{\slashed{\vec{p}}_\genericT}
\newcommand{\mptmag}{\slashed{{p}}_\genericT}

\newcommand{\definmath}[2] {\def#1{\ifmmode#2\else$#2$\fi}}
\newcommand{\invfb}{\ensuremath{{\mathrm{fb}^{-1}}}}
\newcommand{\ourintlumi}{10\,\invfb{}}

\newcommand{\mtstar}{{\ensuremath{m_{\rm T}^{\star}}}}
\newcommand{\mttrue}{{\ensuremath{m_{\rm T}^{\mathrm{true}}}}}
\newcommand{\mtbound}{{\ensuremath{m_{\rm T}^{\mathrm{bound}}}}}

\newcommand{\hideIt}[1]{{}}

\begin{document}   
\title{Re-weighing the evidence for a light Higgs boson in dileptonic
  W-boson decays 
}
\date{\today}

\author{Alan J. Barr}
\email{a.barr@physics.ox.ac.uk}
\affiliation{Denys Wilkinson Building, Keble Road, Oxford, OX1 3RH, United Kingdom}

\author{Ben Gripaios}
\email{gripaios@hep.phy.cam.ac.uk} 
\affiliation{Cavendish Laboratory, Dept of Physics, JJ Thomson Avenue,
Cambridge, CB3 0HE, United Kingdom}

\author{Christopher G. Lester}
\email{lester@hep.phy.cam.ac.uk}
\affiliation{Cavendish Laboratory, Dept of Physics, JJ Thomson Avenue,
Cambridge, CB3 0HE, United Kingdom}

\begin{abstract} 
We reconsider observables for discovering and measuring the mass of a Higgs boson via its di-leptonic decays $h\rightarrow WW^{(*)}\rightarrow \ell\nu\ell\nu$.
We define an observable generalizing the transverse mass that takes
into account the fact that one of the intermediate $W$-bosons is
likely to be on-shell. We compare this new variable with existing ones
and argue that it gives a significant improvement for discovery in the region
$m_h<2m_W$. 
\end{abstract}   
\maketitle 
The LHC has effectively put an upper bound on the mass of the Standard Model
Higgs boson of around 140 GeV. The low mass region that remains is favoured by electroweak precision tests and theoretical
bias (given the existing lower bound of 114 GeV coming from LEP), but is the most difficult to
probe at the LHC, involving challenging final states with low signal cross
section and large backgrounds. One final state that is relevant right
now comes from Higgs decays to $W$ bosons, which in turn decay
leptonically, $W\rightarrow \ell\nu$. Here we propose a new observable to improve the discrimination
between signal from background in this channel.

To motivate the discussion, let us first consider the transverse mass observable
\cite{vanNeerven:1982mz}, originally used in the discovery of the
$W$-boson \cite{Arnison:1983rp}. One way to define
the transverse mass is
as the observable that gives the greatest lower bound on the mass of
the $W$-boson, given the constraints that follow from conservation of
energy-momentum and from the assumption that particles are on-shell
\cite{Barr:2009jv}. The utility of this definition, which seems rather
cumbersome for something as simple as $W$-boson decays, is that it
allows a generalization to any decay process involving missing energy,
once one makes a hypothesis for the decay topology. The resulting
variable is the natural variable for that topology, in the sense that
it encodes all of the information available from kinematic
considerations alone.\footnote{In the case of pair decays in
  supersymmetric theories, this definition picks out
  \cite{Serna:2008zk,Cheng:2008hk,Barr:2009jv} the variable $m_{\genericT2}$
  \cite{Lester:1999tx,Barr:2003rg}; it may also be applied in
  situations with combinatorial ambiguities \cite{Lester:2007fq}, in
  the presence of initial state radiation \cite{Alwall:2009zu}, or in
  situations where one may assume that invisible particles are
  collinear with visible particles
  \cite{Gripaios:2010hv}. Applications to Higgs boson searches include
  \cite{Gross:2009wi,Gross:2009wia,Barr:2009mx,Barr:2011he}.}
It results in a distribution in which signal events pile up below the
mass of the parent resonance, facilitating both discovery of that
resonance and measurement of its mass.

In \cite{Barr:2009mx} (see also \cite{Barger:1987re,Glover:1988fn}), this logic was used to define a new observable
for measuring the mass of a Higgs boson that decays with significant branching fraction to a pair of charged
leptons (which we refer to henceforth as simply leptons) and a pair of neutrinos. The resulting variable has a simple algebraic form, given by
\begin{equation}\label{eq:truedef}
(\mttrue)^2 = m_{\ell \ell}^2 + 2\left(\sqrt{(m_{\ell \ell}^2 +
    \vec{p}_{\ell \ell \genericT}^2) \mptvec^2}  -
  \mptvec \cdot \vec{p}_{\ell \ell \genericT}\right),
\end{equation}
where $\vec{p}_{\ell \ell\genericT}$ and $m_{\ell \ell}$ are the transverse momentum and invariant
mass of the lepton pair and $\mptvec$ is the measured
missing transverse momentum in the event. 
It was shown in \cite{Barr:2009mx} that this variable gave an
improvement over other variables, both for discovery and mass
measurement. The variable was subsequently adopted by the ATLAS
and CMS collaborations, who found a further advantage in the form of a reduced correlation with other variables used for discriminating
signal and backgrounds \cite{ATLAS:2011aa}.

Though the variable defined in \cite{Barr:2009mx} uses all the
kinematic information in the {\em final} state, there remains the
possibility that advances can be made using the internal kinematic
structure of the Higgs decay.  Indeed, one or both of the W-bosons
produced by the Higgs
decay will be almost on-shell \cite{Djouadi:2005gi}, at least for a
Higgs mass in the region where there is significant branching fraction
to $WW^{(*)}$. We will demonstrate that \eqref{eq:truedef} is not, therefore, the optimal observable.

Similar ideas were recently applied in \cite{Barr:2011he} to the decay of a light
Higgs into $\tau \tau$. In that case, both of the $\tau$ mesons will
be on-shell in the interesting region of Higgs masses, and it was
shown that the variable, $\mtbound$, that takes this into account in its definition significantly outperforms the variable $\mttrue$ defined in Ref.~\cite{Barr:2009mx}. 

Let us then reconsider the decay $h\rightarrow WW^{(*)}$ and the dominant background -- continuum $WW$ production. We define a new variable, \mtstar{}, 
defined to give the greatest lower bound on the mass of the Higgs,
subject to the assumption that one or other of the two $W$-bosons is
produced on-shell. Explicitly, denoting the 4-momenta of the charged
leptons and neutrinos by $p_{1,2}^\mu$ and $q_{1,2}^\mu$,
respectively, we minimize $(p_{1}^\mu + p_{2}^\mu + q_{1}^\mu +
q_{2}^\mu)^2$ subject to the constraints that either $(p_{1}^\mu +
q_{1}^\mu)^2 = m_W^2$ or $(p_{2}^\mu +
q_{2}^\mu)^2 = m_W^2$, and $\vec{q}_{1\genericT} + \vec{q}_{2\genericT}=\mptvec$.\footnote{We have not been able to obtain a simple
  algebraic form for $\mtstar{}$; instead we evaluate it via an
  algorithm implemented on a standard personal computer. The relevant code is
  available on-line \cite{code}. Unlike the
  variable $\mtbound$ \cite{Barr:2011he}, one may show that $\mtstar{}$ is well-defined
  for any input momentum configuration.}

It is a simple matter to show that $\mtstar{} \rightarrow \mttrue$ in
the limit that $m_W \rightarrow 0$, so that there is no advantage to be
gained in using $\mtstar{}$ in place of $\mttrue$ if $m_h$ (or
$\hat{s} $ for background events) is large compared to $m_W$. (Proof: Assignments of the unknown neutrinos' momenta that yield a value of $\mttrue$ for the invariant mass of the
$WW$ system correspond to (i) vanishing invariant mass of the di-neutrino
system and (ii) vanishing relative rapidity between the di-lepton and
di-neutrino systems. One of these assignments also yields (iii) vanishing
invariant mass for one lepton-neutrino pair and thus satisfies the
extra constraint that defines $\mtstar{}$ in the limit $m_W=0$. To
wit, (i) and (iii) may be satisfied trivially by assigning vanishing
three-momentum to one of the neutrinos; the three-momentum of the
other neutrino is then fixed uniquely by the measured
$\mptvec$,
which fixes the transverse components, and by (ii), which fixes the
longitudinal component.\footnote{Similar arguments show 
that the $\mtbound$ variable does not coincide with $\mttrue$ in the same
limit:
to enforce the intermediate mass-shell constraints, both neutrinos
would have to be assigned vanishing three-momentum, which would be
incompatible with the observed non-vanishing $\mptvec$.}) Away from
the limit of small $m_W$, and in particular for $m_h < 2m_W$, the
extra constraint requiring one of the $W$-bosons to be on-shell
implies that $\mtstar{} > \mttrue$. Moreover, since $\mttrue < \mtstar{} \leq
m_h$ for signal events, we might hope for an increased number of
signal events, relative to background, in a region below $m_h$,
increasing our ability to discriminate between the background-only and
signal-plus-background hypotheses in an analysis based on counting
events in such a region.

We now compare the performance of the two variables $\mttrue$ and $\mtstar{}$, using a simulation of LHC events corresponding to \ourintlumi{} of integrated luminosity.
We use the {\tt HERWIG}~6.505~\cite{Corcella:2002jc,Marchesini:1991ch} Monte
Carlo generator, with LHC beam conditions ($\sqrt{s}=7~\rm{TeV}$).
Our version of the generator includes the fix to the $h \to WW^{(*)}$
spin correlations described in \cite{herwig-bug-fix}.

We generate unweighted events for Standard Model Higgs boson production ($gg\rightarrow h$)
and for the dominant background, $q \bar{q} \rightarrow WW$.
The detector resolution is simulated by smearing the magnitude of the missing momentum vector 
with a Gaussian resolution function of width
$\sigma_{\mptmag}/\mptmag = 0.4\,{\rm GeV}^{1/2}/\sqrt{\Sigma}$
where $\Sigma$ is the sum of the $|\vec{p}_{\genericT}|$ of all visible fiducial particles.

Selection cuts are applied based on~\cite{Aad:2009wy,ATLAS-CONF-2011-111}, requiring:
\begin{itemize}
\item{Exactly two leptons $\ell\in\{e,\mu\}$ with $p_\genericT > 15$~GeV and $|\eta|<2.5$}
\item{Missing transverse momentum, $\mptmag > 30$~GeV}
\item{ $12~\GeV < m_{\ell\ell} < 300~\GeV$}
\item{No jet with $p_\genericT > 20$~GeV }
\item{$Z\rightarrow \tau\tau$ rejection: the event was rejected if
  $|m_{\tau\tau}-m_{Z}|<25~\GeV$ and 
  $0<x_i<1$ for both $i\in \{1,2\}$\footnote{The variable $x_i$ is the 
  momentum fraction of the $i$th tau carried by its daughter lepton and $m_{\tau\tau}$ is the di-tau invariant mass.
  They are calculated using the approximation that each $\tau$ was collinear with its daughter lepton.}}
\item{Relative azimuth $\Delta\phi_{\ell\ell} < \Delta\phi_{\ell\ell}^{\rm max} $}
\item{Transverse momentum of the $W$ pair system, $p_{\genericT\,WW} > 30$\,GeV}
\item{The appropriate transverse mass (\mtstar{} or \mttrue{}) must satisfy $0.75\times m_h < m_{\rm T} < m_h$}.

\end{itemize}
The value chosen for $\Delta \phi_{\ell\ell}^{\rm max}$ is either 1.3 or 1.8, 
where we use the value for which the larger discovery potential is expected.

\begin{figure}
 \begin{center}
\includegraphics[width=0.9\columnwidth,height=0.5\columnwidth]{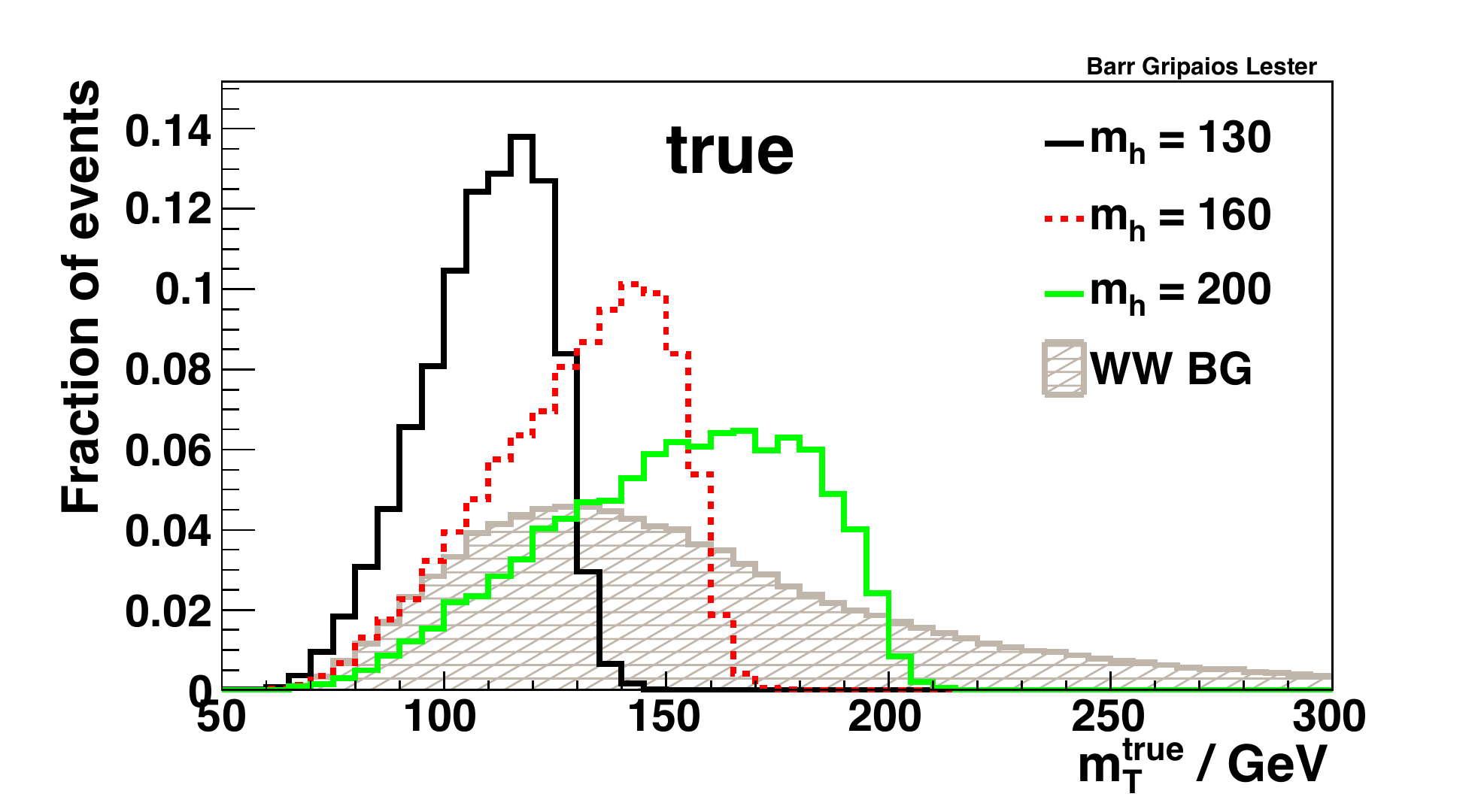}\\
\includegraphics[width=0.9\columnwidth,height=0.5\columnwidth]{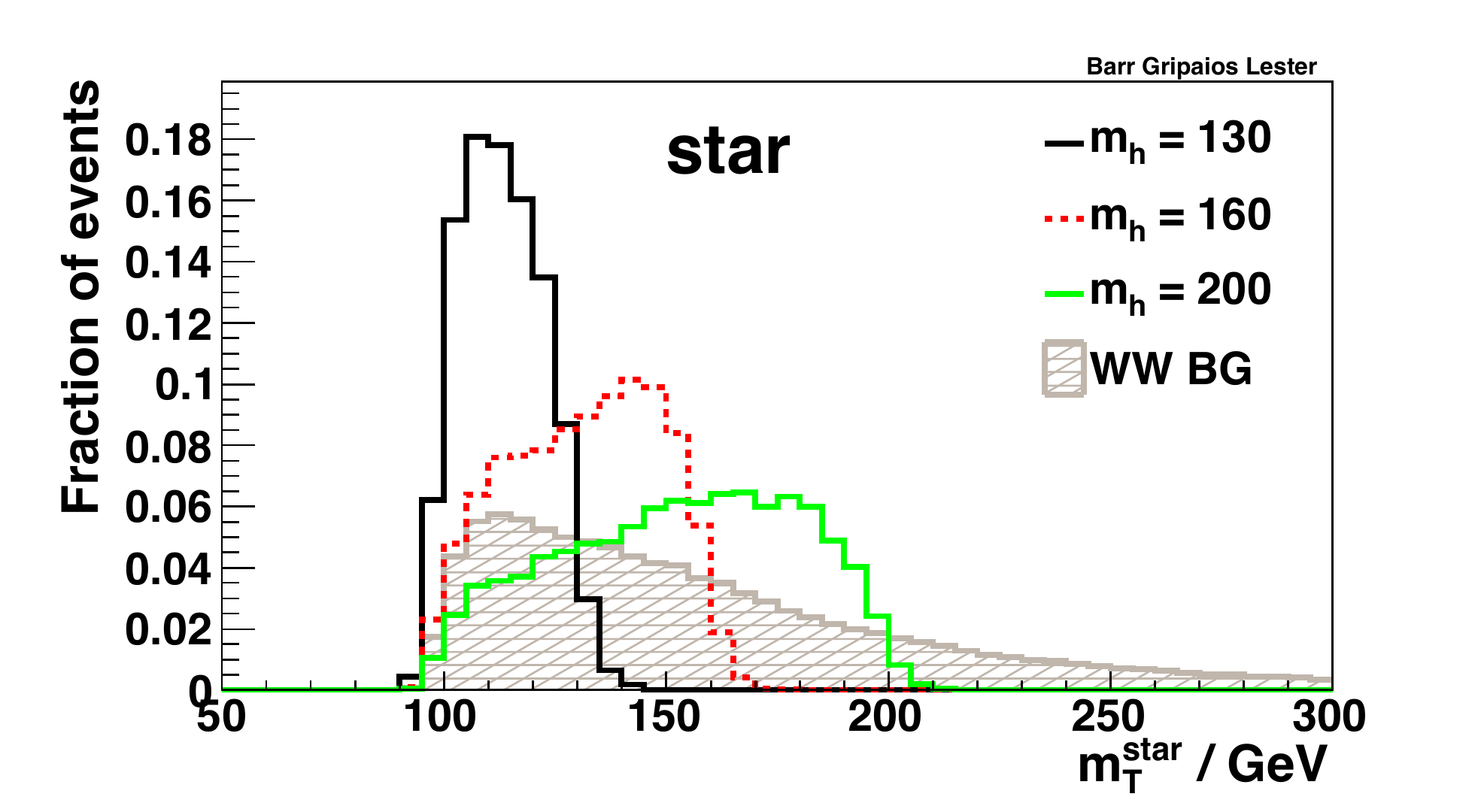}
\includegraphics[width=0.9\columnwidth,height=0.5\columnwidth]{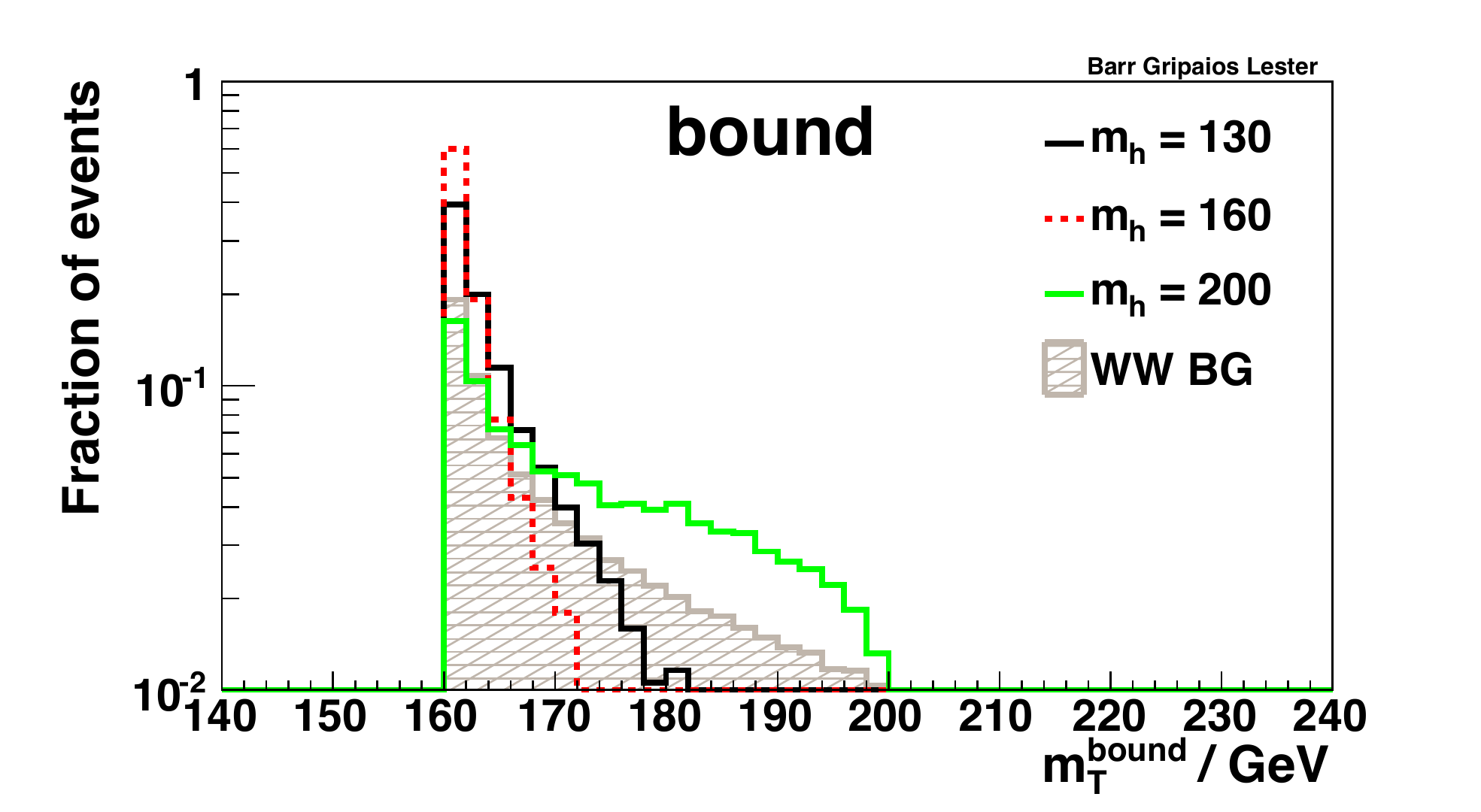}
\caption{
Simulation of $h\rightarrow WW$ signal (for $m_h \in$ $\{130$, 160, 200$\}$\,GeV) for the variables  $\mttrue{}$ (above), $\mtstar{}$ (middle) and $\mtbound$ (below). The shading gives the shape of the dominant $WW$ background. It should be noted that a logarithmic $y$-axis scale (and displaced $x$-axis) has been used when plotting $\mtbound$.
}
\label{fig:1}
\end{center}
\end{figure}

In Fig.~\ref{fig:1} we show sample distributions of the variables $\mttrue$ and $\mtstar{}$ for both the signal (for various $m_h$) and $WW$ background before any cuts are applied. As expected, the signal distributions for both variables are bounded above by $m_h$ (the upper endpoint of the distribution in the absence of resolution and finite width effects).
The $W$ mass-shell constraint means that, by construction, \mtstar{} is larger than $m_W$.
The effect of the additional constraint is that both signal and
background events having $\mttrue<m_W$ must migrate to values larger
than $m_W$, while events with $\mttrue$ significantly larger than
$m_W$ are little changed (i.e. have $\mtstar\approx\mttrue$), in
accordance with the arguments given above. We also show, for comparison, the distribution of the variable $\mtbound$.
This need not be bounded above by $m_h$ if the assumption that both $W$-bosons are on-shell is invalid.

\begin{figure}
 \begin{center}
 \includegraphics[width=0.99\columnwidth,height=0.6\columnwidth]{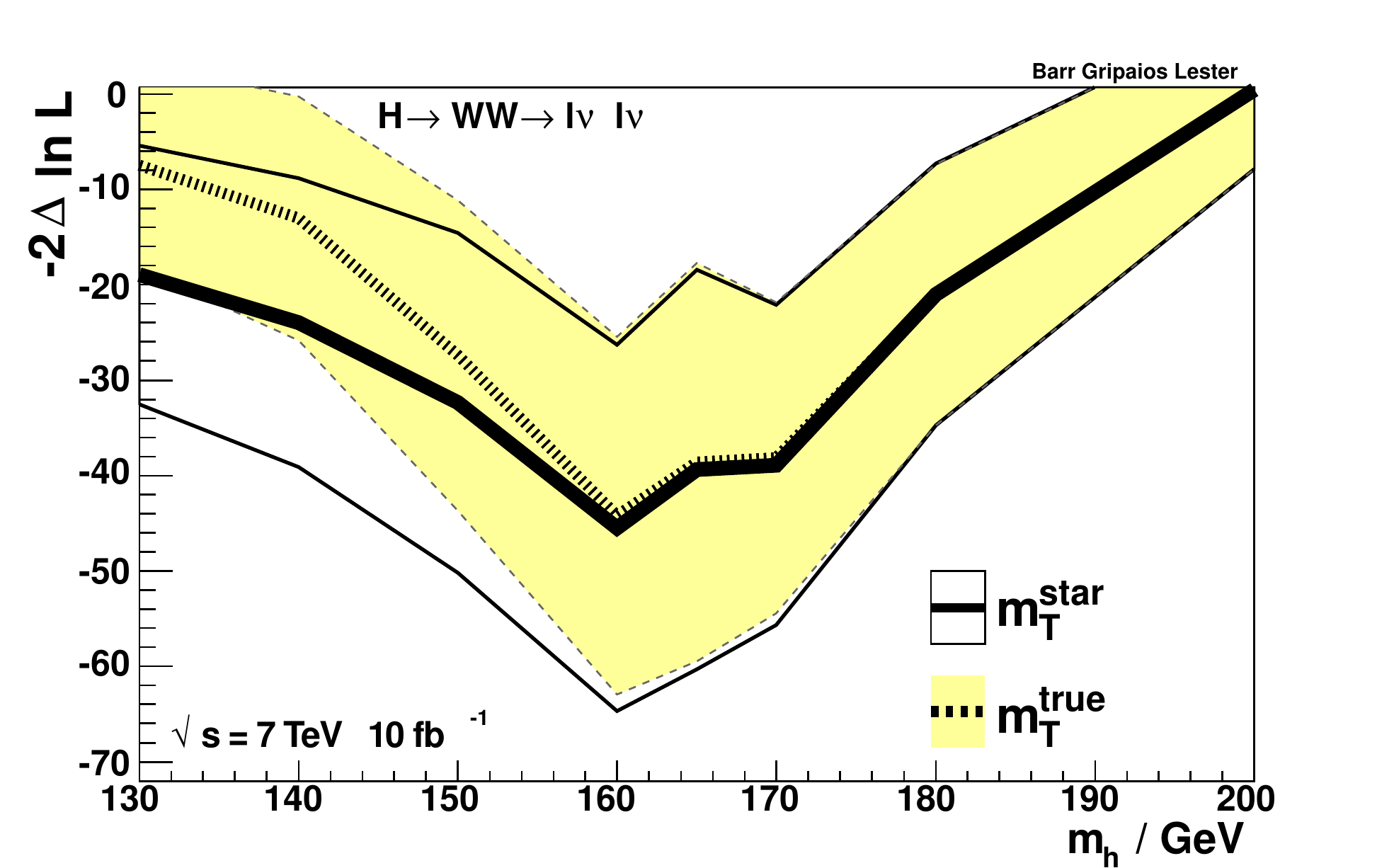}
\caption{
Higgs boson discovery potential as a function of $m_h$, using only $\mttrue$ (dotted, shaded) or $\mtstar{}$ (solid, open).
The center of each band indicates the difference in log likelihood 
between models with and without a Higgs boson contribution.
Lower values correspond to better discovery potential.
The half-width of the each band gives the root-mean-squared over 50 trial samples.
The integrated luminosity simulated is \ourintlumi.
}
\label{fig:2}
\end{center}
\end{figure}

In Fig.~\ref{fig:2} we compare the Higgs discovery potential, as a
function of $m_h$, using only $\mttrue$ (dotted, shaded) or
$\mtstar{}$ (solid, unshaded), as measured by the difference in log likelihood
between models with or without a Higgs boson.  In this figure we see
the key result of this note: the Higgs boson discovery potential is
improved in the region $m_h<2m_W$ and is unchanged elsewhere. This gives a strong indication that $\mtstar{}$ is to be preferred to $\mttrue$ for Higgs discovery purposes.  This improvement may, of
course, be mitigated by an increased correlation of the new variable
with other variables used in the analysis; we leave this for further
investigation.  However, we remark that our cuts were not optimised for \mtstar{}.  Optimisation for \mtstar{} might
lead to improvements in discovery potential for
$m_h<2m_W$ which are greater than those we have shown already.
As expected, for Higgs boson masses above $2m_W$ we have $\mtstar{}
\simeq \mttrue$, and so both variables are equally successful there.
This begs the question: ``Are there better ways of searching for Higgs
bosons in this mass range?''  For these larger masses one might be
prepared to make the hypothesis that {\em both} $W$-bosons are
produced on shell, in which case one could employ the variable
$\mtbound$ proposed in \cite{Barr:2011he}. Further simulations~\cite{Barr:2011si} suggest
that using this variable {\em does} lead to an improvement over
$\mttrue$ and $\mtstar$ for $m_h>2m_W$
so it may be possible to gain improved sensitivity for both large and small Higgs boson masses.
\begin{acknowledgments}
This work was supported by the Science and Technology Facilities Council of the United Kingdom, by the Royal Society, by Merton College, Oxford, and by Peterhouse, Cambridge.
We gratefully acknowledge the hospitality of the Les Houches workshop on ``Physics at TeV Colliders", where part of this work was carried out, and T.J.~Khoo for helpful comments.
\end{acknowledgments}
\bibliography{WWStar}
\begin{figure*}[h]
\begin{center}
\noindent\makebox[\textwidth]{%
\includegraphics{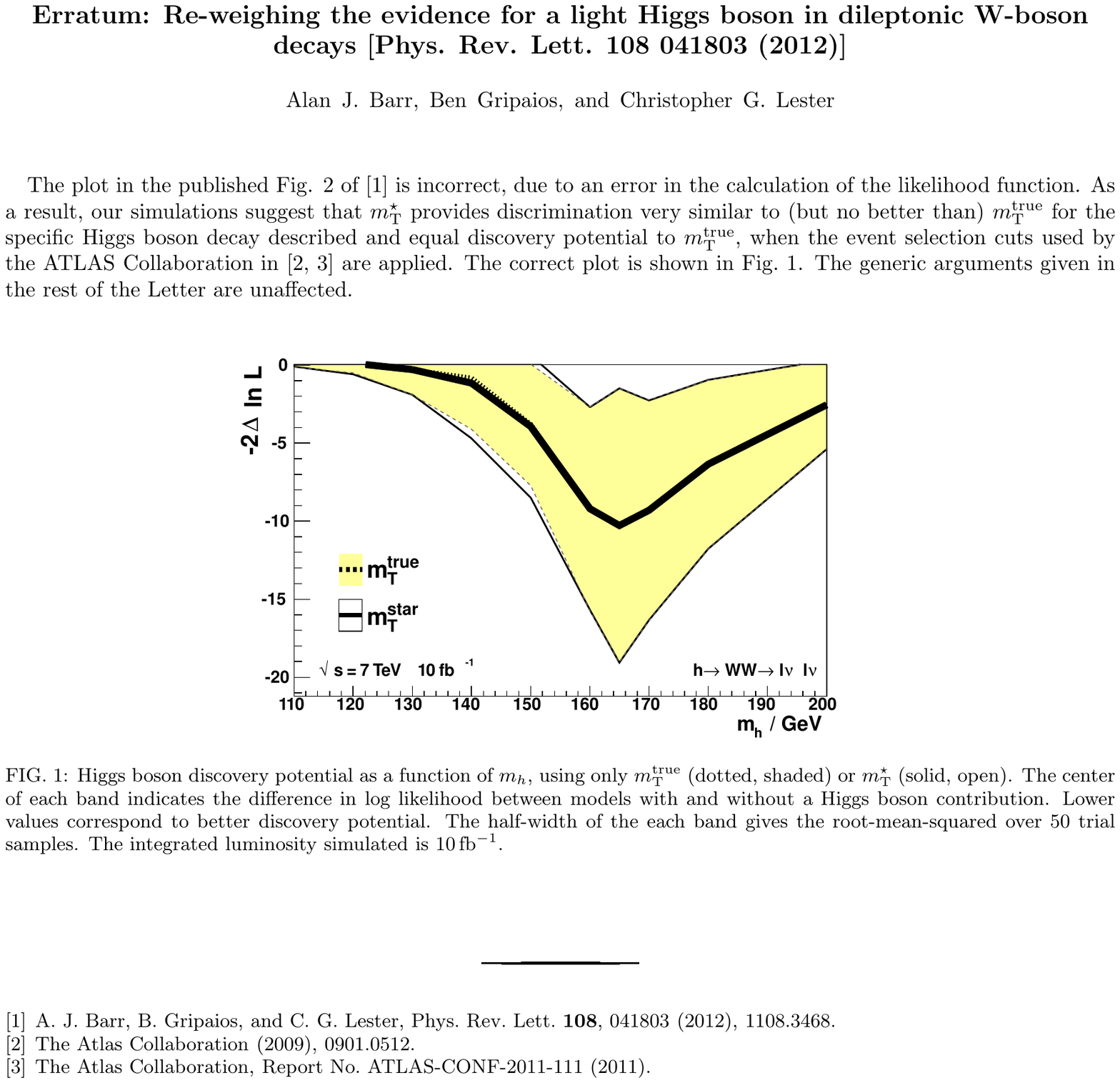}}
\end{center}
\end{figure*}
\end{document}